\documentstyle[epsf,aps,amssymb]{revtex}
\tighten
\begin{document}
\draft
\title{An Exactly Solved Model of Three Dimensional
Surface Growth in the 
Anisotropic KPZ Regime}
\author{M. Pr\"ahofer\thanks{email:praehofer@stat.physik.uni-muenchen.de},
H. Spohn\thanks{email:spohn@stat.physik.uni-muenchen.de}}
\address{Theoretische Physik, Ludwig-Maximilians-Universit\"at,
Theresienstr. 37,\\ D-80333 M\"unchen, Germany}
\date{\today}
\maketitle
\begin{abstract}
We generalize the surface growth model of Gates and Westcott to arbitrary 
inclination. The exact steady growth velocity is of saddle type with 
principal curvatures of opposite sign. According to Wolf this implies
logarithmic height correlations, which we prove by mapping the steady
state of the surface to world lines of free fermions with chiral boundary
conditions.\\[5mm]
\end{abstract}
\newcommand{\ep}{\epsilon}
\newcommand{\D}{\displaystyle}
\newcommand{\T}{\textstyle}
\newcommand{\Ss}{\scriptscriptstyle}
\newcommand{\IE}{\Bbb E}
\newcommand{\IR}{\Bbb R}
\newcommand{\IN}{\Bbb N}
\newcommand{\IZ}{\Bbb Z}
\newcommand{\mod}{\mbox{mod}}
\renewcommand{\span}{\mbox{span}}
\newcommand{\sgn}{\mbox{\tiny sgn}}
\newcommand{\Ord}{{\cal O}}
\newcommand{\y}{\rule{1ex}{1ex}}
\newcommand{\bX}{\mbox{\bf X}}
\newcommand{\by}{\mbox{\bf y}}
\newcommand{\be}{\begin{equation}}
\newcommand{\ee}{\end{equation}}
\newcommand{\bea}{\begin{eqnarray}}
\newcommand{\eea}{\end{eqnarray}}
\newcommand{\ba}{\begin{array}}
\newcommand{\ea}{\end{array}}
\renewcommand{\d}{\mbox{d}}
\section{Introduction}
In its simplest form, deterministic surface growth is governed by
\be\label{detgrowth}
\frac{\partial}{\partial t}h({\bf x},t)=v(\nabla h({\bf x},t)).
\ee
Here $h({\bf x},t)$ is the surface profile at time $t$ relative to a reference 
plane, ${\bf x}\in{\IR}^2$, and (\ref{detgrowth}) just expresses 
that the local 
velocity $v$ along $h$ depends only on the local slope ${\bf u}$, 
${\bf u}=\nabla h$.
Usually microscopic growth mechanisms, like deposition and evaporation of 
atoms, are noisy.
Kardar, Parisi, and Zhang \cite{KPZ} modelled this randomness by adding noise 
and relaxation to (\ref{detgrowth}). Expanding in 
the slope relative to ${\bf u}_o$, they obtained the KPZ equation (in $2+1$
dimensions)
\be\label{KPZequ}
\frac{\partial}{\partial t}h=v({\bf u}_o)+\sum_{i=1}^2\partial_iv({\bf u}_o)
\frac{\partial h}{\partial x_i}+\frac12\sum_{i,j=1}^2\partial_i\partial_j
v({\bf u}_o)\frac{\partial h}{\partial x_i}\frac{\partial h}{\partial x_j}
+\nu\Delta h+\xi
\ee
with $\xi$ space-time white noise. If $\partial_i\partial_jv=\delta_{i,j}
\lambda$, $\lambda\not=0$, then the nonlinearity is of the standard form 
$\frac12\lambda(\nabla h)^2$.
The large space-time scaling behavior is then governed by a
non-Gaussian strong coupling fixed point. We refer to 
\cite{KrugSpohn,Barab,Halpin} as surveys.

Wolf \cite{Wolf} noticed that the Taylor coefficient $\partial^2v$ is really 
a symmetric $2\times 2$ matrix which allows for the possibility 
$\det\partial^2v\leq0$.
Based on a one-loop RG calculation he concluded that in this case the 
nonlinearity
in (\ref{KPZequ}) is irrelevant and the statistics of the growing surface is
identical to the one of an equilibrium interface. In particular, in the steady
state the height-height correlations grow logarithmically with distance.

At the time the prediction of Wolf came as a great surprise. In our
paper we support his claim by a model surface for which we determine the 
growth velocity 
and the steady state correlations explicitly. Our model
was proposed and studied by Gates and Westcott \cite{GW} who, 
in particular, computed
the steady state. Since detailed balance is not available, there is no
general method for obtaining the steady state. It is defined only 
implicitly as the invariant distribution of some stochastic
evolution.
Thus an explicit example is a most welcome exception. Gates and Westcott
considered only tilts of the form ${\bf u}=(u_1,0)$, which does not suffice to
check the Wolf scenario.
To extend their surface model to arbitrary tilt and to obtain its steady
state is straightforward. One merely has to impose chiral boundary 
conditions
for the height lines. Our main advance is to handle such
boundary conditions in the limit of large volume through a transfer matrix
which includes the ``center of mass'' coordinate.
We compute the average velocity $v({\bf u})$ and show that 
$\det\partial^2v\leq0$ for
arbitrary ${\bf u}\neq(0,0)$. To complete the argument we also determine the 
height-height correlations and verify their logarithmic behaviour.
Presumably with some extra effort along the lines of \cite{Spohn} one 
could also show that the statistics is Gaussian on a large scale.

\begin{figure}[ht]
\begin{center}\mbox{\epsfxsize11cm\epsffile{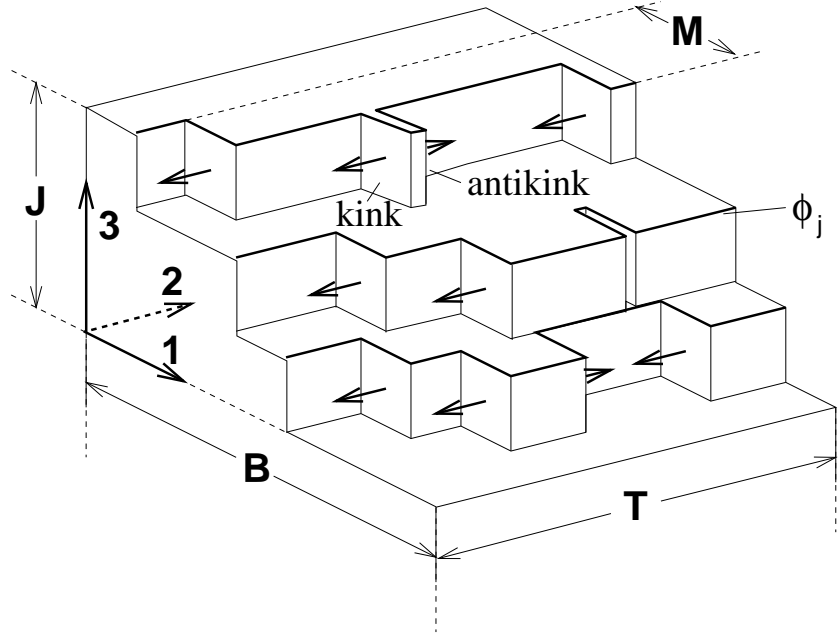}}\end{center}
\caption{\em\label{surface} A piece of the crystalline 
surface. The deterministic motion is indicated.}
\end{figure}

Let us define the Gates-Westcott model of a crystalline surface 
(see Figure \ref{surface}). At a given time it is specified
by the height profile $h(m,\tau)$. Here $m\in\IZ$, $\tau\in\IR$, and $h\in\IZ$.
If also $\tau\in\IZ$, then the bulk crystal would be built up from unit cubes.
Mathematically it is somewhat simpler to allow for real $\tau$. We assume
that $h(m,\tau)$ is non increasing in $m$. Therefore we can introduce the 
collection of height lines $\{\phi_j(\tau)\}$ such that
\be
h(m,\tau)=J-j\quad\mbox{for}\quad\phi_j(\tau)\leq m<\phi_{j+1}(\tau).
\ee
$\phi_j$ and $\phi_{j+1}$ are the boundaries of the terrace 
with height $J-j$. We want to allow for arbitrary step heights along
the discrete $1$-direction. 
Formally, this means that we admit also a terrace width $0$. The height
lines satisfy then the constraint $\phi_j(\tau)\leq\phi_{j+1}(\tau)\in\IZ$
for all $\tau\in\IR$. On the other hand along the continuous $2$-direction
we only admit step height $1$. Thus a height line $\phi_j$ is specified by 
its location, say at $\tau=0$, $\phi_j(0)$, and the location of the rightward
steps (=kinks), $\phi_j(\tau_-)+1=\phi_j(\tau_+)$, and the leftward steps
(=antikinks) , $\phi_j(\tau_-)-1=\phi_j(\tau_+)$ as in Figure \ref{surface}.

Physically we consider a crystal which is in contact with its vapor phase at
a pressure slightly higher than equilibrium. Atoms from the gas phase are 
deposited on the surface. We neglect evaporation. Energetically the fastest
process is deposition at (anti)kink sites. In first approximation this leads
to a constant speed motion of kinks and antikinks along the steps. 
The second fastest
process is deposition at steps away from kink sites, which is modelled as
purely random. Nucleation on the terraces is neglected.
For a single height line the dynamics is known as polynuclear 
growth (PNG) model 
\cite{PNG}. To summarize, the surface dynamics has a stochastic and a 
deterministic piece. 
In the deterministic part every kink moves with velocity $-w$,
$w>0$, and every antikink with velocity $w$ along the $2$-direction. Whenever 
a kink
and antikink pair meet they simply annihilate each other. In addition 
kink-antikink pairs are randomly generated with uniform rate $\xi$ provided
the constraint $\phi_j(\tau)\leq\phi_{j+1}(\tau)$ is not violated. This means,
if allowed, one has $\phi_j(\tau_0)\longrightarrow\phi_j(\tau_0)+1$ at some
$\tau_0$. Under the deterministic motion the newly created kink-antikink pair 
at $(\tau_0,\phi_j(\tau_0))$ immediately separates with speed $2w$. Since
the higher lying terraces expand at the cost of lower lying ones, the 
surface grows along the $3$-direction.

In order to determine the steady state we have to work in finite volume
with the slope ${\bf u}$ imposed by appropriate boundary conditions. This 
will be
done in Section \ref{steady}. In Section \ref{ferm} we map the height lines
to fermionic world lines. Their weight corresponds to noninteracing fermions. 
However there is a complicating constraint due to the $u_2$-tilt of the 
surface.
We compute the free energy and the two point function of the constrained
fermionic system. In Section \ref{results} we relate these results to the 
growth velocity and the height-height correlations.

\section{Steady state}
\label{steady}
We restrict to periodic surface configurations obeying 
$\phi_{j+J}(\tau)=\phi_j(\tau)+B$ and $\phi_j(\tau+T)=\phi_j(\tau)+M$,
where $T\in\IR^+$, $B,J,M\in\IN$. This induces the slope 
${\bf u}=(u_1,u_2)$ with $u_1=-J/B$ and $u_2=-u_1M/T$ 
(see Figure \ref{surface}). For convenience we choose $u_1<0, u_2 >0$. The 
remaining inclinations follow by symmetry.
In order to determine a surface configuration it then suffices to specify $J$ 
height lines $\phi_1(\tau),\ldots,\phi_J(\tau)$
modulo $B$ in the interval $0\leq\tau\leq T$. The height in the $3$-direction
is encoded only modulo $J$.
Let $x_l^{(j)}\in[0,T]$ be the position of the $l$-th kink for height line
$j$, $l=1,\ldots,n^{(j)}+M$ and $y_l^{(j)}\in[0,T]$ be the position of the 
$l$-th antikink, $l=1,\ldots,n^{(j)}$. We set 
${\bf n}=(n^{(1)},\ldots,n^{(J)})$, 
${\bf \|n\|}=2(n^{(1)}+\cdots+n^{(J)})+MJ$, the total number of kinks and 
antikinks, and ${\Phi}_0=(\phi_1(0+),\ldots,\phi_J(0+))$. 
The space of height line configurations $\Gamma$ 
decomposes then as the disjoint union $\Gamma=
\dot{\bigcup}_{{\bf n},\Phi_0}\Gamma({\bf n};\Phi_0)$,
where $\Gamma({\bf n};\Phi_0)$
is some subset of $[0,T]^{{\bf \|n\|}}$ as defined through the constraints 
already explained (we define $[0,T]^0$ to be a single point). 
Furthermore for fixed ${\bf n}$ the 
$\Gamma({\bf n};\Phi_0)$ are glued together in a way that is
determined by the condition that (anti-)kinks leaving the interval $[0,T]$
reappear smoothly on the opposite side, forming the manifold $\Gamma({\bf n})$
(we avoid the subtleties of defining
properly the manifold $\Gamma({\bf n})$).
The stationary measure $P$ restricted to $\Gamma({\bf n})$
is given by 
\be\label{distrib}
P\Big|_{\Gamma({\bf n})}\Big.=
p({\bf \|n\|})\prod_{j=1}^J\bigg(\prod_{l=1}^{n^{(j)}+M}{\d}x_l^{(j)}
\prod_{l=1}^{n^{(j)}}{\d}y_l^{(j)}\bigg)\quad,\quad p(n)=\eta^n/Z.
\ee
Here $\eta=\sqrt{\xi/2w}$ and the normalization
\be
Z=\sum_{n=0}^\infty\eta^n\sum_{{\bf n},\Phi_0;{\bf \|n\|}=n}
|\Gamma({\bf n};\Phi_0)|
\ee
with $|\cdot|$ denoting the ${\bf \|n\|}$-dimensional volume 
(for ${\bf \|n\|}=0$ we set $|\Gamma({\bf n};\Phi_0)|=1$).

To verify the stationarity of $P$ we first note that the Lebesgue measure on 
$\Gamma({\bf n})$ is left invariant by the deterministic part of the dynamics.
At a configuration $\phi$ with a total number of $\|n\|$ kinks and antikinks
there is a gain in probability with rate $2w V_\phi p({{\bf \|n\|}+2})$ through
the annihilation of kink-antikink pairs. Here 
$V_\phi=\sum_{j=1}^J\int_0^T{\d}\tau\theta(\phi_{j}(\tau)-\phi_{j-1}(\tau))$
, $\theta(x)=1$ for $x>0$ and $\theta(x)=0$ otherwise, is the ``number'' of 
configurations immediately before the annihilation event which yields $\phi$.
The loss in probability occurs with
rate $\xi V_\phi p({\bf \|n\|})$ since $V_\phi$ is also  
the ``number'' of 
configurations which arise from $\phi$ by generating a kink-antikink pair.
Equating gain and loss we obtain (\ref{distrib}).
For $\eta\neq0$ the process is ergodic on $\Gamma$ and therefore the 
stationary measure $P$ is unique. (Any configuration can be evolved to a
reference configuration by properly adding kink-antikink pairs in the course
of the deterministic evolution).

The measure $P$ and the partition function $Z$ depend on $T,B$ and $J,M$. 
We want to take the infinite volume limit $B\to\infty$, $T\to\infty$ 
choosing $J,M$ such that the slope ${\bf u}$ is fixed.

\section{Fermi liquid with chiral boundary conditions, free energy}
\label{ferm}
From (\ref{distrib}) we see that, subject to the constraint of not 
crossing, every configuration of height lines has the weight 
$\eta^{{\bf \|n\|}}$. This is precicely
the weight of free fermion world lines in the Euclidian (imaginary time)
set-up \cite{BakVillain}. The surface tilt induces chiral boundary conditions.

\begin{figure}[ht]
\begin{center}\mbox{\epsfxsize9cm\epsffile{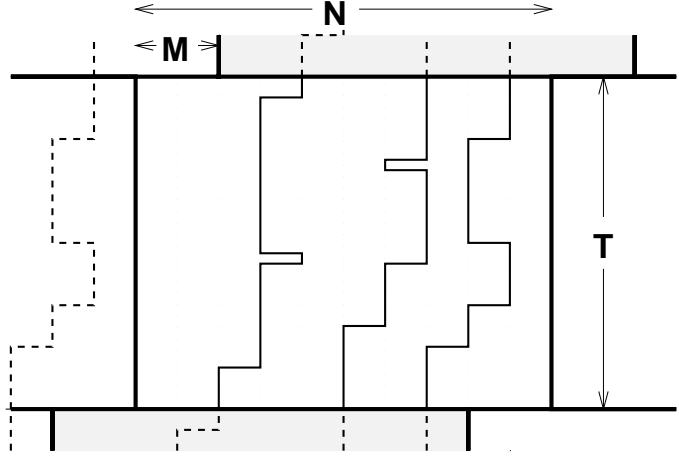}}\end{center}
\caption{\em\label{proj} The mapping of the surface in Figure \ref{surface}
to fermion world lines. Left and right boundaries are identified.
The upper and lower boundaries are identified modulo a relative shift
of $M$ sites.}
\end{figure}

Fermion world lines do not overlap. This can be achieved simply by redefining 
\be
\omega_j(\tau)=\phi_j(\tau)+j-1,\quad j=1,\ldots,J\,\, ,
\ee
with periodized version
\be
\tilde\omega_j(\tau)=\omega_j(\tau)\quad \mod\quad N,\quad N=B+J\,,
\ee
(see Figure \ref{proj}).
We introduce the relative center of mass
\be
m(\tau)=\int_0^\tau{\d}\tau'\sum_{j=1}^J\dot\omega_j(\tau')\quad\in\IZ.
\ee
Clearly $m(\cdot)$ increases by $1$ at every kink and decreases
by $1$ at every antikink. The chiral boundary conditions are then 
\be
m(T)=JM\quad,\quad\tilde\omega_j(T)=\tilde\omega_j(0)+M\quad \mod\quad N, 
\quad j=1,\ldots,J.
\ee

The Fermi liquid lives on $\{1,\ldots,N\}$. Let $\{a^{+}_x,a_x\}$ be
the corresponding Fermi field operators. They act on the fermion Fock
space $\cal F$, which is spanned by $\{a^{+}_{x_1}\ldots a^{+}_{x_l}
|\Omega\rangle\}$ with $|\Omega\rangle$ the Fock vacuum.
Since $m(\tau)\in\IZ$, we introduce the auxiliary Hilbert space $\ell_2$
with the standard orthonormal basis $\{|m\rangle,m\in\IZ\}$, 
$|m\rangle=(\ldots,0,1,0,\ldots)$
at the $m$-th position.
On this space let $D$ be the (unitary) right shift, $D|m\rangle=|m+1\rangle$. 
We define the transfer operator 
\be 
H=-\eta\sum_{x=1}^N(a^{+}_{x+1}a_x\otimes D+a^{+}_x a_{x+1}
\otimes D^{-1}),\quad a_{N+1}=a_1,
\ee
acting on ${\cal F}\otimes \ell_2$. It expresses that a (anti)kink is linked
to the corresponding change in $m(\tau)$.
Let $|x_1,\ldots,x_J;m\rangle=a^{+}_{x_J}\ldots a^{+}_{x_1}|\Omega\rangle
\otimes|m\rangle$ with pairwise distinct $x_j$. Then the matrix element
\be 
\langle y_1,\ldots,y_J;m|e^{-TH}|x_1,\ldots,x_J;0\rangle
\ee
is given by the sum over all fermion world lines with weight $\eta^n$, $n$
the number of fermion jumps, and 
subject to no overlap, such that $\tilde\omega_j(0)=x_j$, $\tilde\omega_j(T)
=y_j$, $j=1,\ldots,J$, and $m(T)=m$. For the partition function $Z$ we need 
$m(T)=JM$ and $y_j=x_j+M\,\,\mod\,\, N$. A fermion configuration is shifted
through the total momentum operator defined by
\be
e^{iP}a_xe^{-iP}=a_{x+1}\quad,\quad a_{N+1}=a_1.
\ee
Then
\be\label{partition}
Z=\langle 0|\mbox{tr}_J\Big[
(e^{iP}\otimes D^J)^{-M} e^{-TH}\Big]|0\rangle
\ee
with $\mbox{tr}_J$ the trace over the Fock subspace containing $J$ particles.

By the same argument we obtain for the density correlations
\be\label{correlations}
\langle(a^{+}_xa_x)(\tau)a^{+}_ya_y\rangle=Z^{-1}\langle0|\mbox{tr}_J\Big[
(e^{iP}\otimes D^J)^{-M}e^{-(T-\tau)H}a^{+}_xa_xe^{-\tau H}a^{+}_ya_y\Big]
|0\rangle. 
\ee
In (\ref{partition}) and (\ref{correlations}) we take the limit $T\to\infty$,
with $M=\alpha T+o(1)\in\IZ$, thereby fixing an average drift $\alpha\in\IR$, 
and subsequently the limit $N\to\infty$ with $J=\rho N+o(1)\in 2{\IN}+1$,  
thus fixing the fermion density $\rho\in[0,1]$. We choose only odd $J$ to 
simplify notation somewhat later on.
We remark that we expect the final result to be independent of the order
of limits. However, choices other than the one discussed here look 
considerably more difficult and we did not persue this issue any further.

Let us proceed with the partition function. $H$, $P$ and $D$ are diagonalized 
through Fourier transform. Let $\Lambda=\{1,\ldots,N\}$. Then the dual is
$\Lambda^\ast=\{-\pi+\frac{2\pi}N,\ldots,\pi-\frac{2\pi}N,\pi\}$. For $\IZ$
the dual is $[-\pi,\pi]$. We set $a_k=N^{-1/2}\sum_x\exp(-ikx)a_x$ for 
$k\in\Lambda^\ast$ and represent $|m\rangle$ by 
$\exp(-imq)\in L^2([-\pi,\pi],(2\pi)^{-1}{\d}q)$. Then
\be
H=-2\eta\sum_{k\in\Lambda^\ast}\cos(k+q)a^{+}_ka_k
\ee
considered as multiplication operator in $q$ on 
$L^2([-\pi,\pi],(2\pi)^{-1}{\d}q)$. Also
\be
e^{iP}\otimes D^J=\exp\Big[-i\sum_{k\in\Lambda^\ast}(k+q)a^{+}_ka_k\Big]
\ee
on the $J$-particle subspace.
Note that $[H,e^{iP}\otimes D^J]=0$.

We have
\be
Z=\int\frac{dq}{2\pi}\sum_{k_1<\cdots<k_J}
\exp\bigg( 2\eta T\sum_{j=1}^J\cos(k_j+q)
+iM\sum_{j=1}^J(k_j+q)\bigg).
\ee
We write $\sum_{j=1}^J\cos(k_j+q)=A_{\{k\}}\cos(\phi_{\{k\}}+q)$, where 
$A_{\{k\}}=|e^{ik_1}+\cdots+e^{ik_J}|$ and 
$\phi_{\{k\}}=\arg(e^{ik_1}+\cdots+e^{ik_J})$,
and use the integral representation of the modified Bessel function,
$I_n(x)=\pi^{-1}\int_0^\pi{\d}q\,e^{x\cos (\phi+q)}\cos(n(\phi+q))$ 
for arbitrary $\phi\in\IR$ and integer $n$,
to obtain
\be\label{exactpartition}
Z=\sum_{k_1<\cdots<k_J}\exp({-iM(k_1+\cdots+k_J-J\phi_{\{k\}})})I_{MJ}
(2\eta TA_{\{k\}}).
\ee
If $A_{\{k\}}$ is maximal, we call $\{k\}$ a ground state mode. By periodicity
we have exactly $N$ of them. The contribution of all
other modes in (\ref{exactpartition}) is exponentially supressed 
for large $T$. Thus
\be 
Z\simeq N\,I_{MJ}(2\eta TA),
\ee 
where $A=|\sum_{\kappa=1}^J\exp(2\pi i\kappa/N)|=N\sin(\pi\!\rho)/\pi+o(1)$ 
for large $N$.
The asymptotic expansion for the modified Bessel function,
\be\label{Besselasympt}
I_n(n/z)=({2\pi n})^{-1/2}({z+\sqrt{1+z^2}})^{-n}\exp({n\sqrt{1+1/z^2}})
\big(1+{\cal O}(n^{-1})\big),
\ee
yields immediately the specific free energy of the 
fermion system in the infinite volume
\be
f=\lim_{N\to\infty}\lim_{T\to\infty}
(TN)^{-1}\ln Z=\alpha\rho\big(\eta_s/\eta_a+\ln\eta-\ln(\eta_s+\eta_a)\big)
\ee
with $\eta_a=\alpha\pi\!\rho/(2\sin\pi\!\rho)=\lim_{T,N\to\infty} MJ/2TA$ 
and $\eta_s=\sqrt{\eta^2+\eta_a^2}$. As explained below, 
$\eta_s\pm\eta_a$
take the role of effective right and left jump rates for the fermions. 
As a further consequence of (\ref{Besselasympt}) we note the relation
\be\label{asympt}
\lim_{n\to\infty}\frac{I_{n-x}(n/z-c)}{I_n(n/z)}=(z+\sqrt{1+z^2})^x
\exp({-c\sqrt{1+z^2}})
\ee 
for later use.

\section{Correlations}

We turn to the expectation values for density, current, and to their 
correlations.
The operator for the density at $x$  is $\rho_x=a^{+}_x a_x$. To obtain
the density at imaginary time $\tau$ we have to evolve $a^{+}_x a_x$with the 
transfer operator $e^{-\tau H}$.
The operator for the rightward fermion current from $x$ to $x+1$ is determined
through the limit
\bea
j^+_x&=&\lim_{t\to0}\frac1t(a_{x+1}^{+} a_{x+1}a_x a_x^{+})(t)
(a_{x}^{+} a_{x}a_{x+1} a_{x+1}^{+})(0)
\nonumber\\
&=&\frac{\partial}{\partial t}(a_{x+1}^{+} a_{x+1}a_x a_x^{+})(t)
(a_{x}^{+} a_{x}a_{x+1} a_{x+1}^{+})(0)\Big|_{t=0}
\nonumber\\
&=&[a_{x+1}^{+} a_{x+1}a_x a_x^{+},H]
a_{x}^{+} a_{x}a_{x+1} a_{x+1}^{+}
\nonumber\\
&=&\eta(a^{+}_{x+1} a_x\otimes D)
\eea
by the fermion anticommutation rules. Analoguously the operator for the 
leftward current from $x$ to $x-1$,
$j^-_x=\eta(a^{+}_{x-1} a_{x}\otimes D^{-1})$.

Let us denote $j_x^{(+1)}=j_x^+$, $j_x^{(0)}=\eta\rho_x$ and 
$j_x^{(-1)}=j_{x}^-$. To determine their expectations, we recall that 
in the limit $T\to\infty$ the trace $\mbox{tr}_J$ reduces to the sum 
$\sum_{\kappa\in\Lambda^\ast}\langle\cdot\rangle_\kappa$, where 
$\langle\cdot\rangle_\kappa$ is the expectation value with respect to 
the $\kappa$-ground state $a^{+}_{\kappa-\pi (J-1)/N}\cdots
a^{+}_{\kappa+\pi (J-1)/N}|\Omega\rangle$. Because of periodicity these
expectation
values are invariant under arbitrary translations in time, space, and even 
in dual space. Thus for $\beta\in\{-1,0,1\}$ we have
\bea
\eta^{-1}\langle j^{(\beta)}_x(\tau)\rangle&&=\langle a^{+}_{\beta} a_{0}
\otimes D^{\beta}\rangle=N^{-1}\sum_{kk'}e^{-ik\beta}
\langle a^{+}_{k} a_{k'}\otimes D^{\beta}\rangle=
\nonumber\\
&&\hspace{-1.5cm}=\lim_{T\to\infty}Z^{-1}N^{-1}\sum_k
\int\frac{{\d}q}{2\pi}\sum_\kappa e^{2\eta TA\cos(\kappa+q)+iMJ(\kappa+q)}
e^{-i(k+q)\beta}\langle a^{+}_k a_k\rangle_\kappa
\nonumber\\
&&\hspace{-1.5cm}=\lim_{T\to\infty}N^{-1}\sum_k\sum_\kappa 
e^{-i(k-\kappa)\beta}
\langle a^{+}_k a_k\rangle_\kappa
Z^{-1}\int\frac{{\d}q}{2\pi}
e^{2\eta TA\cos(\kappa+q)+i(MJ-\beta)(\kappa+q)}
\nonumber\\
&&\hspace{-1.5cm}=\lim_{T\to\infty}N^{-1}\sum_k
e^{-ik\beta}
\langle a^{+}_k a_k\rangle_0
{I_{MJ-\beta}(2\eta TA)}/{I_{MJ}(2\eta TA)}
\nonumber\\
&&\hspace{-1.5cm}=\Big(\alpha J/2\eta A+
\sqrt{1+(\alpha J/2\eta A)^2}\Big)^{\beta}
N^{-1}\sum_k e^{-ik\beta}\langle a^{+}_k a_k\rangle_0.
\eea
In the limit $N\to\infty$ we obtain
\be 
\langle\rho_x(\tau)\rangle=\rho
\ee
and
\be\label{current}
\langle j^{\pm}_x(\tau)\rangle=\frac12\Big(\sqrt{(\alpha\rho)^2
+(2\eta\sin \pi\!\rho/\pi)^2}\pm\alpha\rho\Big).
\ee

We define the structure function
\be
S^{\beta\beta'}(x,\tau)=\eta^{-2}\langle j^{(\beta)}_x(\tau) 
j^{(\beta')}_0(0)\rangle
-\langle j^{(\beta)}_x(\tau)\rangle\langle j^{(\beta')}_0(0)\rangle
\ee
in the infinite volume limit. Assuming $\tau>0$ and using the same method as
above, we obtain for finite $N$
\bea
S^{\beta\beta'}_N(x,\tau)
&=&N^{-2}\sum_{k\kappa k'\kappa'}e^{-i(k-\kappa)x-i(k\beta+k'\beta')}
\langle (a_k^+a_{\kappa}\otimes D^{\beta})(\tau)
(a^+_{k'} a_{\kappa'}\otimes D^{\beta'})(0)\rangle
\nonumber\\
&&-\langle j^{(\beta)}_x(\tau)\rangle\langle j^{(\beta')}_0(0)\rangle
\nonumber\\
&=&N^{-2}\sum_{kk'}e^{-i(k-k')x-i(k\beta+k'\beta')}
\langle (a_{k}^+a_{k'}\otimes D^{\beta})(\tau)
(a^+_{k'} a_{k}\otimes D^{\beta'})(0)\rangle
\nonumber\\
&=&\lim_{T\to\infty}Z^{-1}N^{-2}\sum_{kk'}\sum_\kappa
\int\frac{{\d}l}{2\pi}
e^{-i(k-k')x-i((k+q)\beta+(k'+q)\beta')}\times
\nonumber\\
&&\times e^{2\eta TA\cos(\kappa+q)
-2\eta\tau(\cos(k+q)-\cos(k'+q))+iMJ(\kappa+q)}
\langle a^{+}_k a^{}_{k'}a^+_{k'}a^{}_k\rangle_\kappa
\nonumber\\
&=&\lim_{T\to\infty}Z^{-1}N^{-2}\sum_{kk'}\sum_\kappa
e^{-i(k-k')x-i((k-\kappa)\beta+(k'-\kappa)\beta')}
\langle a^{+}_k a^{}_k\rangle_\kappa\langle a^{}_{k'}a^+_{k'}\rangle_\kappa
\times
\nonumber\\
&&\times\int\frac{{\d}l}{2\pi}
e^{2\eta T(A-\Delta A)\cos(\kappa-\Delta\kappa+q)+i(MJ-\beta-\beta')(\kappa+q)}
\nonumber\\
&=&\lim_{T\to\infty}N^{-3}\sum_\kappa \sum_{kk'}
e^{-i(k-k')x-i(k-\kappa)\beta-i(k'-\kappa)\beta'}
\langle a^{+}_k a^{}_k\rangle_\kappa\langle a^{}_{k'}a^+_{k'}\rangle_\kappa
\times
\nonumber\\
&&\times e^{-i(MJ-\beta-\beta')(\Delta\kappa)}
{I_{MJ-\beta-\beta'}(2\eta T(A-\Delta A))}/{I_{MJ}(2\eta TA)}
\nonumber\\
&=&N^{-2}\sum_{kk'}e^{-i(k-k')x-i(k\beta+k'\beta')}
\langle a^{+}_k a^{}_k\rangle_0\langle a^{}_{k'}a^+_{k'}\rangle_0
(\alpha J/2\eta A+\sqrt{1+(\alpha J/2\eta A)^2})^{\beta+\beta'}\times
\nonumber\\
&&\times 
\exp[{-2\eta\tau\sqrt{1+(\alpha J/2\eta A)^2}(\cos k-\cos k')}{-i\tau\alpha J(\sin k-\sin k')/A}]
\eea
with $\Delta A=(\tau/T)(\cos(k-\kappa)-\cos(k'-\kappa))+
{\cal O}(N^{-1}T^{-2})$, 
$\Delta\kappa=(\tau/{TA})(\sin(k-\kappa)-\sin(k'-\kappa))
+{\cal O}(N^{-2} T^{-2})$. In the last step we used once more 
(\ref{asympt}) and shifted $k$ and $k'$ by $\kappa$.
Taking the limit $N\to\infty$ results in
\bea
S^{\beta\beta'}(x,\tau)&=&((\eta_a+\eta_s)/\eta)^{\beta+\beta'}
\int_{-\pi\!\rho}^{\pi\!\rho}\frac{{\d}k}{2\pi}
\int_{\pi\!\rho}^{2\pi-\pi\!\rho}\frac{{\d}k'}{2\pi}
\nonumber\\
&&\hspace{-1.5cm}
\exp[{-2\tau\big(\eta_s(\cos k-\cos k')+i\eta_a(\sin k-\sin k')\big)
-i(k-k')x-i(k\beta+k'\beta')}]
\eea
for $\tau>0$ and $S^{\beta\beta'}(x,\tau)=S^{\beta\beta'}(-x,-\tau)$ by 
symmetry.
The Fourier transform of $S^{\beta\beta'}$ with respect to $x$ equals
\bea
\hat S^{\beta\beta'}(K,\tau)&=&
(({\eta_a+\eta_s})/\eta)^{\beta+\beta'}\times
\nonumber\\
&&\hspace{-2cm}\times\int_{|(K/2)-\pi\!\rho|}^{(K/2)+\pi\!\rho}
\frac{{\d}k}{2\pi}
\exp[{-4\tau\sin(K/2)(\eta_s\sin k-i\eta_a\cos k)-ik(\beta+\beta')
-iK(\beta-\beta')/2}]
\eea
for $K\geq0$ and 
$\hat S^{\beta\beta'}(K,\tau)=\hat S^{\beta\beta'}(-K,\tau)^\ast$ 
for $K<0$. For small $|K|$ we have 
$\hat S^{\beta\beta'}_K(\tau)\propto|K|+{\cal O}(K^2)$,
and therefore the correlations decay as $\hat S^{\beta\beta'}_{x}(\tau)
\sim x^{-2}$
for large $x$ and constant $\tau$.
We remark that the same expectation values and two-point correlations are
obtained for a pure Fermi liquid with the usual periodic boundary conditions 
but with a non-hermitian transfer operator 
$H=\sum_x(\eta_s+\eta_a)a_{x+1}^+a_x+(\eta_s-\eta_a)a_{x}^+a_{x+1}$.

\section{Surface fluctuations}\label{results}
\begin{figure}[ht]
\begin{center}\mbox{\epsfxsize9cm\epsffile{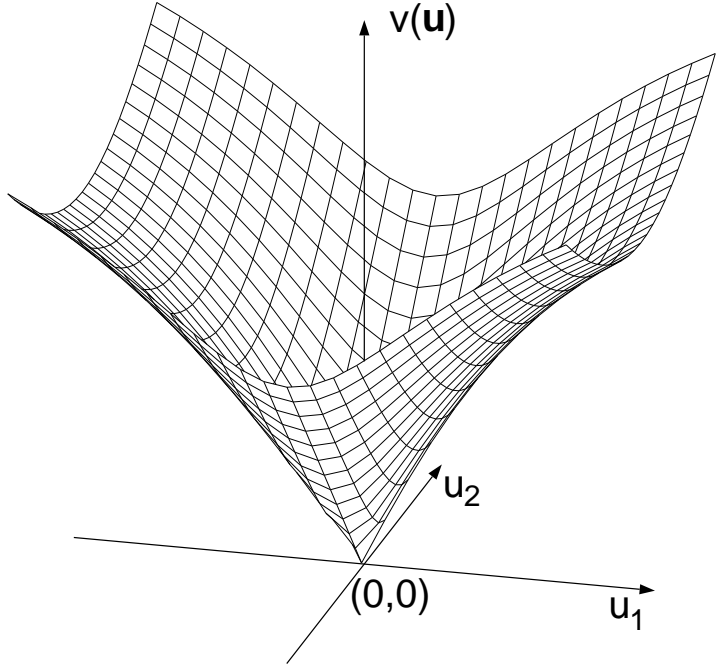}}\end{center}
\caption{\em\label{growth} The growth velocity $v$ as a function of the 
surface tilt ${\bf u}$.}
\end{figure}

The growth velocity $v$ of the surface in the $3$-direction is proportional to 
$n$, the total number of kinks and antikinks in $\{1,\ldots,B\}\times[0,T]$,
\be
v={w}({BT})^{-1}\langle n\rangle.
\ee
In the fermion picture $n$ equals the integrated, total leftward 
and rightward fermion current, i.e. 
$n=\sum_{x=1}^N\int_0^T{\d}\tau(j^+_x(\tau)+j^-_x(\tau))$. Thus, using
(\ref{current}),
\bea\label{growthvelo}
v&=&w(B+J)B^{-1}\sqrt{(2\eta\sin\pi\!\rho/\pi)^2+(\alpha\rho)^2}
\nonumber\\
&=&w\frac{1+|u_1|}\pi\sqrt{\Big(2\eta\sin\frac{\pi|u_1|}{1+|u_1|}\Big)^2+
\Big(\frac{\pi|u_2|}{1+|u_1|}\Big)^2},
\eea
because the surface inclination ${\bf u}$ is related to the fermion density 
$\rho$ and drift $\alpha$ as $\rho=|u_1|/(1+|u_1|)$ and $ \alpha=|u_2/u_1|$.
By symmetry (\ref{growthvelo}) holds for arbitrarty ${\bf u}\in {\IR}^{2}$. 
As can be seen from Figure \ref{growth}, for small ${\bf u}$ we have 
qualitatively $v(u)\simeq|{\bf u|}$, which is characteristic for models 
ignoring nucleation
and island formation on terraces. Except for ${\bf u}=(0,0)$ the growth 
velocity is saddlelike, i.e. 
$\det\partial^2v\leq0$. Thus the RG analysis of Wolf 
applies with the prediction that the
Gates-Westcott crystal surface is in the Edwards-Wilkinson universality class.
If so, the surface 
fluctuations should increase logarithmically for large separation.

To check this property for the Gates-Westcott model we parametrize the surface
relative to the new plane obtained by rotating the 
$1$-$2$-plane by $45^\circ$ around the $2$-axis. The new height function 
$\tilde h$ is then given implicitly for each $\tau$ by 
$\tilde h(x,\tau)=h(m,\tau)-m$
where $x=h(m,\tau)+m$. $\tilde h$ is more explicitly related to the fermion
world lines $\omega_x(\tau)$, since
\be
\tilde h(x,\tau)=\tilde h(0,\tau)+\sum_{y=1}^x\omega_y(\tau).
\ee
The height difference
between $(0,0)$ and $(x,t)$, $\Delta(x,t)=\tilde h(x,t)-\tilde h(0,0)$,
equals the number of fermion lines passing between $(0,0)$ and $(x,t)$. We 
denote the number of fermions at time $t$ in $\{1,\ldots,x\}$ as 
$N_{[1,x]}(t)$, $N_{[y,x]}(t)=\sum_{x'=y}^x\rho_{x'}(t)$, and the integrated 
fermion current from site $x$ to site $x+1$ during the time interval $[0,t]$ as
$J_{[0,t]}(x)$,
$J_{[0,t]}(x)=\int_0^t{\d}\tau (j^+_x(\tau)-j^-_x(\tau))$. The height 
difference $\Delta(x,t)$ is then expressed as
\be
\Delta(x,t)=N_{[1,x]}(0)-J_{[0,t]}(x).
\ee
The mean height difference is given by
\be
\langle \Delta(x,t)\rangle=\rho(x-\alpha t).
\ee
To compute the height-height correlations we use fermion particle conservation
\be
J_{[0,t]}(x)-J_{[0,t]}(y)=N_{[y+1,x]}(0)-N_{[y+1,x]}(t)
\ee
for $x>y$. Then, using that
current and density correlations decouple for large 
distances, we have
\bea
\langle\Delta(x,t)^2\rangle&-&\langle \Delta(x,t)\rangle^2=
\nonumber\\
&=&\langle (N_{[1,x]}(0)-J_{[0,t]}(x))^2\rangle-(\rho x-\alpha\rho t)^2
\nonumber\\
&=&\lim_{y\to\infty}\langle (N_{[1,x]}(0)-J_{[0,t]}(x)+J_{[0,t]}(-y))\times
\nonumber\\
&&\times(N_{[1,x]}(0)-J_{[0,t]}(x)+J_{[0,t]}(y))
\rangle-\rho^2 x^2
\nonumber\\
&=&\lim_{y\to\infty}\langle (N_{[-y+1,x]}(t)
-N_{[-y+1,0]}(0))(N_{[1,y]}(0)-N_{[x+1,y]}(t))\rangle
-\rho^2x^2
\nonumber\\
&=&\lim_{y\to\infty}\int\frac{{\d}k}{2\pi}\frac{e^{ik(y-1)}-e^{ik(x+1)}}
{1-e^{-ik}}\Big(\frac{e^{ik}-e^{ik(y+1)}}{1-e^{ik}}\hat S(k,t)-
\nonumber\\
&&-
\frac{e^{-ik(x+1)}-e^{ik(y+1)}}{1-e^{ik}}\hat S(k,0)\Big)
-\frac{e^{ik(y-1)}-e^{-ik}}{1-e^{-ik}}\times
\nonumber\\
&&\times\Big(\frac{e^{ik}-e^{ik(y+1)}}{1-e^{ik}}\hat S(k,0)-
\frac{e^{-ik(x+1)}-e^{ik(y+1)}}{1-e^{ik}}\hat S(k,-t)\Big)
\nonumber\\
&=&\int\frac{{\d}k}{2\pi}\frac1{1-\cos{k}}(\hat S(k,0)-e^{-ikx}\hat S(k,t)),
\eea
with $\hat S=\hat S^{00}$. The rapidly oscillating terms vanish in the 
limit $y\to\infty$, since they are multiplied by continuous $2\pi$-periodic
functions. For large $|t|$, keeping $x/t$ constant, the leading contribution 
is 
\bea
\langle\Delta(x,t)^2\rangle&-&\langle \Delta(x,t)\rangle^2=
\nonumber\\
&=&\pi^{-2}\int_0^{\pi|t|}\frac{{\d}k}k\Big(1-e^{-2k\eta_s\sin\pi\!\rho}
\cos k(2\eta_a\cos\pi\!\rho-x/t)\Big)+\Ord(1)
\nonumber\\
&=&\pi^{-2}\ln({|t|})+\Ord(1).
\eea
If $|x|$ gets large while $t=o(x)$ we have
\bea
\langle\Delta(x,t)^2\rangle&-&\langle \Delta(x,t)\rangle^2=
\nonumber\\
&=&\pi^{-2}\int_0^{\pi|x|}\frac{{\d}k}k\Big(1-e^{-2k|t/x|\eta_s\sin\pi\!\rho}
\cos k(2(t/x)\eta_a\cos\pi\!\rho-1)\Big)+\Ord(1)
\nonumber\\
&=&\pi^{-2}\ln({|x|})+\Ord(1).
\eea
In conclusion we have shown that indeed the height-height correlations 
increase
logarithmically with separation. Properly speaking the fluctuations
should be considered in the same reference frame as the growth velocity.
However, this will not change their logarithmic growth.

\section*{Acknowledgements}
 We thank J.~Krug for insisting on arbitrary tilts.

\end{document}